\documentstyle[manuscript,aps]{revtex}
\title{Comparison of simulation
methods for 
charged systems of slab geometry \footnote{We dedicate the article to
our colleague J P Badiali}.}
\author{M Mazars, J-M Caillol, J-J Weis and D Levesque}

\address{Laboratoire de Physique Th{\'e}orique, UMR 8627,
  B{\^a}t. 210, Universit{\'e} de Paris-Sud \\ 91405 Orsay Cedex,
  France}
\def\bb#1{\mbox{\boldmath $#1$}}
\def\bm#1{\mbox{\boldmath $#1$}}
\def\t2{\mbox{  }}
\def\rst1{\mbox{ }}

\begin{document}
\maketitle

\begin{abstract}
Using the specific model of a system of like charged ions confined
between two planar like charged surfaces, we compare the predictions for
the energy and density profile of four simulation methods available
to treat the long range Coulomb interactions in systems periodic in
two directions but bound in the third one. 
Monte Carlo simulations demonstrate unambiguously complete agreement
between the results obtained with these methods where the potential
between charges is solution of Poisson's equation in the simulation
cell with adequate boundary conditions. The practical advantages of the
different methods is assessed.
\keywords Simulations, Monte Carlo, Coulomb interaction, colloids
\pacs 61.20.Ja,68.15.+e,82.70.-Dd
\end{abstract}

\section{Introduction}
The standard way to treat long range interactions (Coulomb, dipolar,
Yukawa near the Coulomb limit) in simulations  of bulk systems is to
replicate the basic simulation cell periodically in all three directions
of space and apply an Ewald summation technique (EW3D)
\cite{Alle,Sali}. However, in many 
situations of electrochemical, biological or technological interest,
as, for instance, electrolyte solutions between charged surfaces,
charged lipid bilayers in water, suspensions of colloids between glass
plates, clays \cite{Isra,Evan,Grun}, magnetic thin films \cite{Bell},
Wigner crystals \cite{Weis}
etc.., the 
system is finite in one direction thus necessitating a
modification of the convential Ewald method. 
For systems periodic in two dimensions but bounded along the third one
an Ewald summation method (EW2D) for electrostatic
interactions has first been given by Parry \cite{Parr,Parr1} and was later
rederived by various authors \cite{Heye,Leeu,Cich,Harri,Grzy}. 
Unfortunately, the practical use of the EW2D sum is hampered by the
occurrence, in the reciprocal space term, of a double sum over 
different particles which, due to
the complicated way the bounded coordinates enter the
expression, cannot be reduced, as for the EW3D case, to a
sum of order N, a circumstance which renders the method computationally
expensive.
Not surprisingly, only few calculations using EW2D have been reported to
date mainly to test the validity of more approximate  approaches \cite{Spho,Yeh}.
These calculations involve tabulation and inteerpolation of the potentials on a
three-dimensional grid.

Various ``time saving'' schemes have been proposed to bypass the
computational burden of the EW2D method.
The purpose of this paper  is to compare some of these methods for the
specific case of a system of charged particles confined between two charged
planar surfaces separated by a distance of the order of several
particle diameters.
The methods that have been chosen are those which satisfy exactly the
laws of electrostatics (perfect screening, Green's theorem etc..) and do
not present numerical instabilities difficult to control. Thus the 
charged sheet method proposed by Torrie 
and Valleau \cite{Torr,Vall} and its modification by Boda {\em et al}
\cite{Boda} have been discarded.
Indeed the distribution of the point charges of the ions outside the
simulation cell, located in the image cells of the latter, is
represented approximately by a set of uniformely charged planar sheets.
An other method, proposed by Lekner  \cite{Lekn,Lekn1} is based
on rewriting the sum of  forces acting on an ion by a second ion
and its periodic images 
as an absolutely converging infinite sum. 
The energy is obtained by integration of the force. 
A shortcoming of the method is the need, to estimate the sum with a
given precision, to retain a number of terms depending
strongly on the relative position of pairs of particles. 
Because of this technical reason, emphasized in more detail in ref
\cite{Maza}, the method was not included in this investigation.

The four methods we have considered, the first three of which have
already appeared in the literature, are EW3D \cite{Spho,Yeh,Shel,Croz},
a method developed by   
Hautman and Klein \cite{Haut} (HK), the method of hyperspheres (HSG)
\cite{Cail} and that
of concentric spheres (CS) described below. 

A way to treat the long range Coulomb interactions is to place the
slab at the centre of a parallelepipedic simulation box which has
dimension perpendicular to the slab much larger than the width of the
latter and apply the EW3D method. In this way one effectively simulates
an infinite number of parallel slabs. Provided the region of empty space
separating the slabs is sufficiently large one expects the influence of
the periodic images on the behaviour of the system to become negligible.

If the distance between the surfaces confining the particles is small
compared to the extension of the system along the directions parallel to
the surfaces  (narrow slab), the lateral distance $s$ between pairs of
particles will be much larger than the distance $z$ normal to the slab
surfaces and it will be  appropriate to expand the Coulomb pair interaction in
powers of $z/s$ and apply an Ewald sum to the in-plane
component of the interaction. Such an approach was followed by Hautman
and Klein (HK) \cite{Haut}.

An attractive way to investigate the behaviour of Coulomb particles
between charged surfaces which circumvents the cumbersome Ewald sums is
to use a closed space, e.\ g.\ a hypersphere in four-dimensional
Euclidean space \cite{Cail,Pell,Cail1}.
Not only is the electrostatic potential, solution of Poisson's equation
on the hypersphere obtainable in simple closed form \cite{Pell}, thus
avoiding approximation of the interaction potential (as for instance in
EW3D due to truncation of the direct and reciprocal space sums), also
the number of operations necessary to calculate the distances between
particles is reduced with respect to Euclidean space with concomitant
speed up of the simulation procedure.
This geometry has been applied previously to the study
of the attraction between two like charged surfaces neutralized by
solvated counterions in an endeavour to comprehend the stability of charged
lamellar materials such as clays and cement \cite{Pell,Delv} and to the study of the
effective interaction of charged colloidal particles confined between
like charged plates \cite{Alla}.

A system of charged particles occupying the region between the outer and inner
surfaces of two concentric spheres is also a suitable arrangement to
describe, in the limit of sufficiently large radii of the spheres, the
system of charged particles confined between two planar surfaces. 
This geometry has the  advantage that the interaction between charges
and between charges and surfaces is the usual Coulomb potential; however
it may require use of a large number of particles to render the effect of
curvature of the confining surfaces small.

The purpose of this paper is to check, by means of Monte Carlo
simulation,  the agreement between the results of the four
aforementioned methods  for the energy and density 
profile of the system described above  consisting of $N$ like-charged ions between
parallel  like-charged surfaces separated by a distance $h$.
The ions, modelled as hard spheres of diameter $d$ bear a
charge $q$ at their centre and each surface, of area $S$, a uniform
charge density  $\sigma$.
Such a system can be viewed as a crude
model for lamellar liquid crystals formed by ionic amphiphiles
\cite{Guld} or charged lamellar materials like clays or cement
\cite{Pell}. 
The convergence of the results to their thermodynamic limit is estimated
by performing simulations with an increasing number of particles
keeping the distance between surfaces fixed. The speed of this
convergence is obviously an important criterion for  appraisal of  the
relative practical interest of the four methods.

Expressions for the energy calculated with the  different methods
will be given in Sect.\ II together with details of the Monte Carlo
simulations. 
Results for the energy and density profiles obtained with the four
methods are compared in Sect.\ III and conclusions drawn in Sect.\ IV.

\protect\section{Energy expressions}

\protect\subsection{EW3D}
In this method a square slab of ions of thickness $h$
is placed at the centre of a simulation box having dimension $L_z$
normal to the surfaces much
larger  then the lateral dimension $L$  and the system
is extended 
periodically in the three directions of space.
The slab surfaces are located at $z= \pm h/2$ perpendicular to the $z$-axis.
The closest approaches of the ions to the impenetrable surfaces are
therefore $z= \pm \frac{1}{2} (h-d)$.

 To evaluate the total energy of the system
which is the sum of the contributions from the ion-ion $U_{ii}$,
ion-surface $U_{is}$ and
surface-surface $U_{ss}$ interactions it will be convenient to associate to the
ions a uniform neutralizing background (filling the whole simulation
cell) of density $-Nq/V$
and to the charged walls a uniform background of density
$-2 \sigma S/V = -2 \sigma/L_z$
where $V=L_zL^2=L_z S$ is the volume of the simulation cell.
Because of electroneutrality of the system  $Nq + 2\sigma S = 0$, the
backgrounds cancel out 
exactly and will not contribute to the total energy.

The contribution of the ions and their background to the total energy is
given by

\begin{eqnarray}
\label{I2}
U_{ii}  =& & \frac{q^2}{2}  \biggl\{  \sum_{i,j}^{N}
  \sum_{\bb \nu} {}^{'} 
  \frac{{\rm erfc} (\alpha | {\bb r}_{ij} + {\bb \nu}|)}
       {| {\bb r}_{ij} + {\bb \nu}|}  \nonumber \\ 
& &  + \frac{1}{V}  \sum_{i,j  \atop j \neq 
i}^{N}
\sum_{{\bb G} \neq 0} \displaystyle 
 \frac{4 {\pi}^2}{{\bb G}^2} e^{-  \displaystyle {\bb G}^2/{4
  {\alpha}^2}} \  e^{i {\bb G} \cdot ({\bb
  r}_{j} -{\bb r}_{i})}  \nonumber \\
& & -\displaystyle \frac{1}{V}  \sum_{i,j  \atop
  j \neq i}^{N}
  \displaystyle \frac{\pi}{\alpha^2} \biggr\} - \frac{1}{2} C_w   
\end{eqnarray}
In these equations ${\bb r}_{ij} = {\bb r}_{j}- {\bb r}_i$ where ${\bb
r}_{i}$ and ${\bb r}_{j}$ are the
positions of particles $i$ and $j$.

The term  $- \frac{1}{2} C_w $ 
is the self-energy given by
\begin{eqnarray}
\label{I3}
- \frac{1}{2}C_w =&&-\frac{1}{2} q^2 N \biggl \{ - \sum_{\bb \nu \neq 0} 
  \frac{{\rm erfc} (\alpha | {\bb \nu} |)}
       {| {\bb \nu}|} 
  - \frac{1}{V} 
\sum_{{\bb G} \neq 0} \displaystyle 
 \frac{4 \pi^2}{{\bb G}^2} e^{- \displaystyle {\bb G}^2/4 \alpha^2}  \nonumber \\
&& + \displaystyle \frac{2  \alpha}{\sqrt \pi} +
  \displaystyle \frac{\pi}{{\alpha}^2} \biggr\} 
\end{eqnarray}
In Eq.\ (\ref{I2}) ${\bb \nu}$ is a vector of components
($Ln_x$,$Ln_y$,$L_z n_z$) ($n_x,n_y,n_z$  integers) and the prime in the
sum over ${\bb \nu}$ means  that 
the terms $i=j$ must be omitted when ${\bb \nu}=0$. 
The wave-vectors ${\bb G}$ which enter the reciprocal space
contributions to the energy have components
$2 \pi n_x/L$, $2 \pi n_y/L$ and $2 \pi n_z/L_z$

In our calculations the sum on lattice vectors extends over all 
${\bb G}$
subject to $|n_x| \leq 6$,  $|n_y|\leq 6$, $|n_z|\leq 12$ and $ |{\bb n}| \leq n_{max} =12$.
The
$\alpha$ parameter which governs the rate of convergence of the
real- and reciprocal-space contributions was taken sufficiently 
large so that only the terms with ${\bb \nu}=0$ had to be retained in 
Eqs.\ (\ref{I2}) and (\ref{I3}) 

The electrostatic energy between the ions and the two charged walls
including their compensating backgrounds is given by
\begin{equation}
U_{is} = \sum_{i=1}^{N} q V_s(z_i) -\displaystyle \frac{Nq}{V} \int_V
 d {\bb r} V_s(z)
\label{uis}
\end{equation} 
where $V_s(z)$ is the electric potential associated with the electric
field $E_s$ generated by the two charged
surfaces and theit backgrounds.
Due to the symmetry of the simulation cell, this electric field  is
normal to the charged surfaces and depends
only on the $z$-coordinate. Furthermore, as a
consequence of the periodic boundary conditions,
it must satisfy $E_s(-L_z/2)=E_s(L_z/2)=0$.
Applying Gauss's theorem (cf.\ Fig.\ 1) one finds
\begin{eqnarray}
E_s(z) = \left\{ \begin{array}{lrll}
\displaystyle - \frac{8 \pi \sigma}{L_z} z -4 \pi \sigma \quad & -L_z/2 &\leq z&
\leq -h/2 \\
\displaystyle - \frac{8 \pi \sigma}{L_z} z \quad & -h/2 &\leq z &\leq h/2 \\
\displaystyle - \frac{8 \pi \sigma}{L_z} z +4 \pi \sigma \quad & h/2&
\leq z& \leq L_z/2 
\end{array} \right.
\label{Ez}
\end{eqnarray}
and, by integration, for the potential, choosing $V_s(0)=0$ (any additional constant would leave $U_{is}$ unchanged),
\begin{eqnarray}
V_s(z) = \left\{ \begin{array}{lrll}
\displaystyle - \frac{4 \pi \sigma}{L_z} z^2 +4 \pi \sigma z + 2 \pi
\sigma h \quad  & -L_z/2& \leq z& \leq -h/2 \\
\displaystyle - \frac{4 \pi \sigma}{L_z} z^2 \quad & -h/2 &\leq z& \leq h/2 \\
\displaystyle - \frac{4 \pi \sigma}{L_z} z^2 -4 \pi \sigma z + 2 \pi
\sigma h \quad & h/2 &\leq z& \leq L_z/2
\end{array} \right.
\label{Vs}
\end{eqnarray}
In  the periodic system the potential created by the charged surfaces is
thus parabolic. 
The potential $V_s(z)$ can also be obtained by a direct integration of
the Ewald potential over the two charged planes resulting in a
one-dimensional Fourier series which can be summed explicitly to yield 
Eq.\ \ref{Vs}.
Integration of $V_s$ over the volume of the simulation box leads to 
\begin{equation}
U_{is} = \displaystyle - \frac{4 \pi \sigma q}{L_z} \sum_{i=1}^{N} z_i^2   \displaystyle -\frac{Nq \pi \sigma}{3L_z}(3 h^2 -6 L_z h +2 L_z^2)
\end{equation}
Finally, the interaction between the two charged surfaces (including their
background) is
\begin{equation}
U_{ss} = \displaystyle \frac{1}{8 \pi} \int_V d{\bb r} E_s^2(z)
 = \displaystyle \frac{2 \pi S \sigma^2}{3 L_z^2} [h^3 + (L_z-h)^3].
\end{equation}
%
%
\subsection{Hautmann-Klein method.}

In this method the slab, having the characteristics described above, has 
periodic boundary conditions only in the $x$- and $y$- directions.
The origin of coordinates being at the centre of the parallelepipedic
cell, the energy of the system is given by
\begin{equation}
\begin{array}{ll}
\displaystyle U&\displaystyle
=\frac{q^{2}}{2}\sum_{i,j=1}^{N}\sum_{\bm{\nu}}{}^{'}\frac{1}{\mid\bm{r}_{ij}+\bm{\nu}\mid}
+\displaystyle \sigma q \sum_{i=1}^{N} \sum_{\alpha=1}^{2} \int_{S}dx_{\alpha}dy_{\alpha} \sum_{\bm{\nu}}\frac{1}{\mid\bm{r}_i-\bm{r}_{\alpha}+\bm{\nu}\mid}\\
&\displaystyle+\frac{1}{2} \sigma^2 \sum_{\alpha,\beta =
1}^{2} \int_{S}dx_{\alpha}'dy_{\alpha}'\int_{S}dx_{\beta}dy_{\beta} \sum_{\bm{\nu}}\frac{1}{\mid\bm{r}_{\alpha}'-\bm{r}_{\beta}+\bm{\nu}\mid}
\end{array}
\end{equation}
where ${\bb r}_{ij}={\bb r_j}-{\bb r_i}$, ${\bb r_i}$,${\bb r_j}$ coordinates of
the particles $i$ and $j$, and  the vectors ${\bb r}_{\alpha}$ and ${\bb
r}_{\beta}$ 
are the position vectors of points on the surfaces $S_{\alpha}$ ($\alpha
=1,2$) having constant $z$  coordinate $z=h/2$ if $\alpha$ or $\beta =1$
and $z=-h/2$ if $\alpha$ or $\beta =2$.
The sum over ${\bb \nu}$ runs over all vectors of components $(n_xL,n_y
L)$ ($n_x$, $n_y$ integers) perpendicular to the $z$-direction.
The three sums in the expression for $U$ correspond to the interactions
between the particles, the interaction between the particles and the
surfaces $S_1$ and $S_2$ and to that between the two surfaces    $S_1$
and $S_2$, respectively. To each of these sums, due to the infinite
number of replicated cells, is associated  a divergent contribution.
 However, if electroneutrality is taken into account these
divergent terms will cancel.

The evaluation of 
the ion-ion interaction starts with the identity \cite{Haut}
\begin{equation}
\displaystyle \frac{1}{r}= {\Bigl (}
\frac{1}{r}-\sum_{n=0}^{M}a_{n}\frac{z^{2n}}{s^{2n+1}} {\Bigr )}  +\sum_{n=0}^{M}a_{n}\frac{z^{2n}}{s^{2n+1}}
\end{equation}
with
\begin{center}
$\displaystyle a_{n}=\frac{(-1)^{n}(2n)!}{2^{2n}(n!)^{2}}$.
\end{center}
The added and subtracted term represents the first $M+1$ terms in the
binomial expansion of $1/r$ in powers of $z/s$ where $s$ is the
component of $r$ in the plane of the surface.
By introducing a convergence function $h_n(s;\alpha)$ for each term
$1/s^{2n+1}$ the energy can be divided in a short range part
\begin{equation}
U_{ii}^s \displaystyle =
\frac{1}{2}\sum_{i=1}^{N}\sum_{j=1}^{N}q_{i}q_{j}\mbox{\Large (}\sum
_{\bm{\nu}} {}^{'}\frac{1}{r_{ij,\bm{\nu}}}-\sum_{n=0}^{M}a_{n}z_{ij}^{2n}\frac{h_{n}(s_{ij,\bm{\nu}};\alpha)}{s_{ij,\bm{\nu}}^{2n+1}}\mbox{\Large
)} 
\end{equation}
and a long range part
\begin{equation}
U_{ii}^l \displaystyle =
\frac{1}{2}\sum_{i=1}^{N}\sum_{j=1}^{N}q_{i}q_{j}\sum_{n=0}^{M}a_{n}z_{ij}^{2n}\mbox{\Large  
(}\sum_{\bm{\nu}} {}^{'}\frac{h_{n}(s_{ij,\bm{\nu}};\alpha)}{s_{ij,\bm{\nu}}^{2n+1}}\mbox{\Large
)} 
\end{equation}
which can be evaluated in reciprocal space.
In these equations $s_{ij,\bm{\nu}}=|{\bb s_j}-{\bb s_i} + {\bb \nu}|$.
With the choice
\begin{equation}
h_{0}(s;\alpha)=\mbox{erf}(\alpha s)
\end{equation}
and
\begin{equation}
\displaystyle\frac{h_{n}(s;\alpha)}{s^{2n+1}}=\frac{1}{a_{n}(2n)!}\nabla^{2n}\mbox{\large (}\frac{h_{0}(s;\alpha)}{s}\mbox{\large )}
\end{equation}
as proposed by Hautman and Klein
 the
short and  long ranged parts of the electrostatic energy take the form
 \cite{Haut} 
\begin{equation}
U_{ii}^{s} \displaystyle = \frac{1}{2}\sum_{i=1}^{N}\sum_{j\neq
i}^{N}q_{i}q_{j}\mbox{\Large (}\frac{1}{r_{ij}}-\frac{\mbox{erf}(\alpha
s_{ij})}{s_{ij}}-\sum_{n=1}^{M}\frac{1}{(2n)!}z_{ij}^{2n}\nabla^{2n}\mbox{\large
(}\frac{\mbox{erf}(\alpha s_{ij})}{s_{ij}}\mbox{\large )}\mbox{\Large )}
\label{us}
\end{equation}
where only the $\bm{\nu}=0$ term has been retained (see below)
and 
\begin{equation}
\begin{array}{ll}
U_{ii}^{l} &\displaystyle =\frac{\pi}{A}\sum_{i=1}^{N}\sum_{j=1}^{N}q_{i}q_{j}\sum_{n=0}^{M}\frac{1}{(2n)!}z_{ij}^{2n}\sum_{\bm{G}\neq\bm{0}}G^{2n-1}\mbox{erfc}(G/2\alpha)e^{i\bm{G}.\bm{s}_{ij}}\\
&\\
&\displaystyle -\frac{\alpha}{\sqrt{\pi}}\sum_{i=1}^{N}q_{i}^{2}-\frac{\sqrt{\pi}}{\alpha A}\mbox{\large (}\sum_{i=1}^{N}q_{i}\mbox{\large )}^2
\end{array}
\label{ul}
\end{equation}
with ${\bb G}$ a two-dimensional vector of components $2 \pi (n_x/L,n_y/L)$. 
In our simulations we kept terms up to $M=3$ which allowed the short
range part $U_{ii}^{s}$ to be limited to its ${\bb \nu}=0$ term even for
relatively large values of $h$.
Expressions for $h_n(s;\alpha)$ up to M=3 are
\begin{equation}
\left\{\begin{array}{lll}
&h_1(s;\alpha)&\displaystyle=\mbox{erf}(\alpha s)-\frac{2\alpha s}{\sqrt{\pi}}e^{-\alpha^2 s^2}(1+2\alpha^2 s^2)\\
&&\\
&h_2(s;\alpha)&\displaystyle=\mbox{erf}(\alpha s)-\frac{2\alpha s}{\sqrt{\pi}}e^{-\alpha^2 s^2}(1+\frac{2}{3}\alpha^2 s^2-\frac{4}{9}\alpha^4 s^4+\frac{8}{9}\alpha^6 s^6)\\
&&\\
&h_3(s;\alpha)&\displaystyle=\mbox{erf}(\alpha s)-\frac{2\alpha s}{\sqrt{\pi}}e^{-\alpha^2 s^2}(1+\frac{2}{3}\alpha^2 s^2+\frac{4}{15}\alpha^4 s^4+\frac{8}{25}\alpha^6 s^6\\
&&\\
&&\displaystyle\t2\t2\t2\t2 -\frac{112}{225}\alpha^8 s^8+\frac{32}{225}\alpha^{10} s^{10}).
\end{array}
\right.
\end{equation}

The attractive feature of the method is that,
by writing the $z_{ij}^{2n}$ explicitly as polynomials in $z_i$ and $z_j$,
$U_{ii}^l$ becomes a sum of terms each of which involvs products of
two functions having general form
\begin{equation}
F_p(G) =\sum_i^N q_i z_i^p e^{i\bm{G}.\bm{s}_{i}}.
\end{equation}
In the evaluation of $U_{ii}^l$ sum on the pairs of particles is
replaced by sums on particles which, obviously, makes the computation faster.
The contributions to the energy of the interactions between the ions and
the surfaces and between the surfaces are easily obtained 
using the method of  de Leeuw and Perram \cite{Leeu}
and the identity
\begin{equation}
\int_{S}dx_{\alpha} dy_{\alpha} \sum_{\bm{\nu}}f(\mid\bm{r}_{a}-\bm{r}+\bm{\nu}\mid)\equiv\int_{-\infty}^{+\infty}dx_{\alpha}\int_{-\infty}^{+\infty}dy_{\alpha}\rst1 f(\mid\bm{r}_{a}-\bm{r}\mid).
\end{equation}
The divergent terms of these two contributions having been eliminated
through use of the electroneutrality condition, the ion-surface and
surface-surface interactions contribute to the energy $U$ by a constant
term equal to $2 \pi \sigma^2 V$.
The expression for the total energy is therefore given by
\begin{equation}
U = U_{ii}^{s} + U_{ii}^{l} + 2 \pi \sigma^2 V.
\end{equation}
\subsection{Hyperspherical geometry}
In this method the Monte Carlo simulations are performed on a
hypersphere $S_3$ in four-dimensional Euclidian space. Two surfaces of
angular colatitudes $\theta_N$ and $\theta_S = \pi -\theta_N$ separated by
a distance $h=R(\pi-2\theta_N)$ are located symmetrically on opposite
sides of the equator (see Fig.\ 3 of ref.\ \cite{Pell}). $N$ neutralizing
ions of charge $q$
are confined between the two surfaces which bear each a charge density
$\sigma$. For given $h$ and $\sigma$ the aperture $\theta_N$ is obtained
by solving the equation 
\begin{equation}
\displaystyle \frac{h \sin \theta_N}{\pi-2 \theta_N} = {\Biggl ( -
\displaystyle \frac{qN}{8 \pi \sigma} \Biggr )}^{1/2}. 
\end{equation}

Similar to the EW3D case described above, neutralizing backgrounds are
associated to both the ions and the charged surfaces. As a detailed
derivation of the different contributions to the potentiel energy,
ion-ion $U_{ii}$, ion-surface $U_{is}$ and surface-surface $U_{ss}$ can
be found in ref.\ \cite{Pell} only the final expressions are given
here 

\begin{eqnarray}
U_{ii} & = &\displaystyle \frac{q^2}{2 \pi R} \sum_{i=1}^N \sum_{j \ne i}^N
[ (\pi - \theta_{ij}) \mbox{cot} \theta_{ij} -0.5] \nonumber \\ 
&  &  + \displaystyle \frac{N q^2}{2 \pi R} [(\pi - 2\alpha)
\mbox{cot}\alpha +0.5 - \pi/\mbox{sin}\alpha] \\
U_{is} & = &\displaystyle \frac{q q_s}{\pi R} 
[ (\pi - 2\theta_{i}) \mbox{cot} \theta_{i} -1]
 + \displaystyle \frac{4 q_s^2}{ \pi R} (\theta_N \mbox{cot}
\theta_N -1) \\
U_{ss} & = &\displaystyle \frac{ q_s^2}{\pi R} [1 + (\pi -4 \theta_N)
\mbox{cot} \theta_N]. 
\end{eqnarray}
Here the angle $\alpha$ is equal to the ion radius $d/2$ divided by the
hypersphere radius $R$, $\theta_{ij}$ is the angular separation
between particles $i$ and $j$ and $q_s=\sigma S$ ($S$ area of each surface).

\subsection{Concentric spheres}

In the last method we have explored,  the ions occupied  the
region of volume $V$ between two concentric spheres of radii $r_l$ and $r_m = h+ r_l$. The external surface of the inner sphere, of area $S_l=4
\pi r_l^2$,
and the internal surface of the outer sphere, of area $S_m=4 \pi r_m^2$,
bear a charge density $\sigma$. The two surfaces being impenetrable the
distances of closest approach of an ion to the surfaces are $r_l +d/2$
and $r_m-d/2$. The electroneutrality conditions reads
\begin{equation}
\sigma (S_l+S_m) + Nq =0.
\end{equation}
The total energy of the system is readily obtained as
\begin{equation}
U = \displaystyle \frac{1}{2} \sum_{i=1}^N \sum_{j \ne i}^N \frac{q^2}{|
{\bb r}_i - {\bb r}_j|} + \sum_{i=1}^N \displaystyle \frac{S_l \sigma
q}{r_i} 
+  \displaystyle q  N \frac{\sigma S_m}{r_m}
+  \displaystyle   \frac{\sigma^2 S_m S_l}{r_m}
+  \displaystyle  \frac{1}{2} \frac{\sigma^2 S_m^2}{r_m}
+  \displaystyle  \frac{1}{2} \frac{\sigma^2 S_l^2}{r_l}.
\end{equation}   
where the three first terms in the r.h.s.  represent the ion-ion energy, the energy
of the ions with the charged surface $S_l$ and with the surface $S_m$
which creates in the volume $V$ a constant potential. The three last
terms correspond to the energies
associated with the surface charges of $S_l$ and $S_m$.

\section{Results}
In our comparisons we did not consider realistic values of
$q$ and $\sigma$ as for instance envisaged in ref.\ \cite{Pell,Delv}. Our
aim being methodological, we chose values of $q$ and $\sigma$ allowing
for a compelling test of the efficiency  of the four methods to predict
reliably the properties of the confined charged system. In the
following  charges
$q$ and $\sigma$ will be expressed by using reduced units of charge and
distance $(\beta q^2/d)^{1/2}$
and  $d$, respectively.
In these units the surface charge $\sigma$ has been fixed to -1 in all
our simulations and the value of $q$ varied from 0.5 to 5. The distance
between the surfaces limiting the system has been taken to be $h=5$;
however, in order to test the limit of validity of the HK method for
large $h$ a few simulations were performed with $h=8$ and $h=12$.
For given values of $h$ and $\sigma$ electroneutrality entails that the
density of the system increases for decreasing values of $q$ if the
volume of the simulation cell remains unchanged.

The simulations were carried out in the canonical ensemble. Of the order
of $10^5$ MC steps per particle were performed to calculate the energy
and the density profiles when $N \le 3000$. In the CS geometry, for
$ N \approx 15000$, $10^4$ trial configurations per particle were
generated to calculate these quantities. The statistical error on the
energies is of the order of $\pm 0.02$ in units of $kT$. For the two
planar geometries, EW3D and HK, the density profiles are estimated with
an error of $ \sim 0.5 \%$; the statistical error, except for systematic
error due to
curvature effects, is of the same order for the HSG method. In the case 
of the CS geometry, for $N \approx 15000$, the error on the energies is
of the order of $\pm 0.03$ and on the density profiles of the order of
$2-3 \%$. 

In the calculations using the EW3D method, the simulation volume has a
square section of side $L$ and an elongation along the $z$-axis of $L_z$
varying between 60 and 90. For $h=5$ and $L$ between 15 and 30, the
influence of the value of $L_z$ on the simulation results turned out to
be always negligible. Also, within this geometry, if the system confined
between the surfaces has a net dipole moment, a correction to the energy
proportional to the square of the dipole moment normal to the surfaces
can be taken into account to remove the interaction between net dipoles
in the periodically repeated slabs \cite{Yeh}. As the dipole moment in
our system was  small such a correction was not considered.

A systematic comparison of the energies for $0.75 \le q \le 5$ is made in
Table 1. It shows without ambiguity the convergence of the results of
the four methods. It appears that the thermodynamic limit for $U$ is
obtained for $q \ge 1$ as soon as the number of particles is of the
order of 500 for EW3D and HK and of the order of 2000-3000 for the HSG
and CS method. However, at $q \approx 0.75$, i.\ e.\ at densities larger
than 0.4 at fixed $h$ and $\sigma$, at least 10000 particles are
necessary for the energy calculated with CS to agree
within $1 \%$  with the other methods. This result is obviously related
to the important curvature effects expected for radii $r_l$ and $r_m \le
10$.
Figure  \ref{fig2} shows, for the four methods, the variation of $U$
when the ionic charge $q$ varies from 0.5 to 5. A maximum is observed
for $q \approx 0.8$. This maximum occurs as a consequence of the
lowering of $q$ and saturation of the particles in
contact with the confining surfaces due to steric effects. Indeed layers
of particles 
parallel to the surfaces form at high density as evidenced by  
Fig.\ \ref{fig3} which presents the density profiles for the different values
of $q$. For large $q$, layers of highly localized particles are in
contact with the surfaces. When the density increases (i.\ e.\ $q$
decreases)  layers separated by a distance of order $d$ appear in the
volume.
The profiles shown in Fig. \ref{fig3} are obtained by the method of HK and
are indistinguishable from those given by EW3D.

Figures \ref{fig4} and \ref{fig5} allow to compare the density profiles obtained with
the methods EW3D and HK with those evaluated with the methods of
hypersphere and concentric spheres. For $q=2$ excellent agreement is
found for the four methods taking into account the statistical errors
, especially those for the CS method. At $q=0.75$ differences are
manifest and demonstrate the persistence of curvature effects; they are
small in the hypersphere geometry but remain important in the CS
geometry in spite of the large value $N=14036$. This result at $q=0.75$
is in agreement with the slow convergence, in the latter geometry, of
the energy to its thermodynamic limit when $q<1$.

Extrapolations towards $z=\pm 2$ give the values $\rho_c$ of the
profiles when the particles are in contact with the surfaces. It has
been shown that for planar surfaces separated by a distance much larger 
than the diameter $d$ of the particles 
the pressure is given  by \cite{Hend,Carn}
\begin{equation}
P/kT = \rho_c - E_c^2/8 \pi kT
\end{equation}
where $E_c$, the value of the electric field at the surface, equals $4
\pi \sigma$ for planar surfaces.
For the confined systems considered in this work this expression is not
strictly valid anymore. Nonetheless, considering it as an approximation
and using the values of $\rho_c$ given in Table 2, an estimate of $P$
can be obtained. The value of $P$ is small and positive for $q>1$ and
rises to a value of 5 at $q=0.5$ where hard core interactions are
important. The nearly zero value of the pressure for $q>1$ is clearly in
accord with the absence of particles in the central part of the
simulation cell (cf.\ Fig.\ \ref{fig2}) and complete screening of the
charged surfaces by the particles in their vicinity implying an almost
vanishing electric field in this same region of simulation volume.

 Table 2 and Fig.\ \ref{fig6} comparing simulation results for the
 energy and density profiles for  $h=8$
 and $12$ obtained with the EW3D and HK methods, show that the latter
 method remains accurate even for $h/L \sim 0.5$. However, reliable results
 are obtained only if, in Eqs.\ \ref{us} and \ref{ul} for the energy, $M$
 is chosen to be 3.     
        
\section{Conclusion}
The convergence of the results obtained with the four simulation methods
considered is clearly the main conclusion of this study. It
demonstrates the  strict equivalence of the various possible routes to
take into account the long range of the Coulomb interactions : use  
of periodic boundary conditions, a closed system without boundary or
simply a large system with boundaries.
The equivalence is made realized only by the use of potentials
preserving the laws of electrostatics and thus solution of Poisson's
equation associated with the simulation cell and its boundary
conditions.
One can remark that the system under investigation is particularly challenging
to establish this equivalence  because particles
all bear a 
charge of the same sign and screening effects therefore result only from
interactions between the particles and the surfaces.

Depending on the type of system which is simulated, the advantage of
using one method or the other may differ. This work clearly shows that
the EW3D and HK methods apply more favourably to the study of systems
confined by planar surfaces, the thermodynamic limit being reached for
$N \sim 500$. However, with regard to computing time, the method of
hypersphere is the most efficient despite the necessity to use systems
with larger values of $N$ to make curvature effects
negligible. Obviously, the CS method is unfavourable to  study 
planar interfaces as curvature effects get small only for
$N>10000-20000$.
It has been used here mainly to establish without ambiguity the
equivalence between the Ewald and hypersphere ``Coulomb'' potentials and
the usual Coulomb potential.

The calculations carried out with the CS method nevertheless show that
the ionic density profiles are affected in the vicinity of a charged
interface by the curvature of the latter. Such modifications of
the density profiles near planar charged interfaces may be
present in solutions of inverted micelles or 
suspensions of charged colloids of small diameter. 

The determination of the relative efficiency of the simulation methods
for the calculation of the properties of confined systems will be
complete only when pressure, free energy and surface tension will have
been evaluated. Calculation of these quantities is in progress.

\newpage

\begin{figure}
\caption{Simulation setup for EW3D. The slab of width $h$ is placed at
the center of a simulation cell having dimension $L_z$ in the direction normal
to the slab larger than the lateral dimensions $L$. Periodic boundary
conditions are applied in all three directions. To evaluate the electric
field $E_s(z)$ created by the charged surfaces and their backgrounds,
Gauss's theorem is applied to the dotted volume.}
\label{fig1}
\end{figure}
\begin{figure}
\caption{Variation with ionic charge $q$ of the total energy of a system
of like ions confined between planar surfaces bearing a charge density
$\sigma=-1$ and separated by a distance $h=5$, for the four
different simulation methods considered.}
\label{fig2}
\end{figure}
\begin{figure}
\caption{Variation with ionic charge $q$ of the density profile
$\rho(z)$ of  a system
of like ions confined between planar surfaces bearing a charge density
$\sigma=-1$ and separated by a distance $h=5$. The results shown
correspond to the method of HK and are  indistinguishable from those
obtained for EW3D.}
\label{fig3}
\end{figure}
\begin{figure}
\caption{Ion density profile for a slab of width $h=5$, surface charge
density $\sigma=-1$ and ionic charge $q=2$. Comparison between all four
simulation methods. The profile is shown for $z/d>1$.}
\label{fig4}
\end{figure}
\begin{figure}
\caption{Ion density profile for a slab of width $h=5$, surface charge
density $\sigma=-1$ and ionic charge $q=0.75$. Comparison between all four
simulation methods. The profile is shown for $z/d>-0.5$.}
\label{fig5}
\end{figure}
\begin{figure}
\caption{Comparison of the density profiles obtained with the HK and EW3D
methods for (from top to bottom) $h=5$, $\sigma=-1$; $h=8$, 
$\sigma=-1$ and $h=12$, $\sigma=-2$. All results are for $q=1$. For
clarity the profiles 
have been shifted by 0.5 with respect to each other.} 
\label{fig6}
\end{figure}

\newpage          

\begin{table}
\caption{Variation with ionic charge $q$ of the total energy of a system
of like ions confined between planar surfaces bearing a charge density
$\sigma=-1$ and separated by a distance $h=5$, for the four
different simulation methods considered. The number $N$ of ions is
given in brackets, $\beta=1/kT$, $T$ temperature.}

\vspace{1cm}
\begin{center}
\begin{tabular}{|ccccc|}
\hline
\hline
&&&&\\    
$q$ & $\beta U_{EW3D}/N$ &$\beta U_{HK}/N$ & $\beta U_{HSG}/N$ & $\beta U_{CS}/N$ \\
&&&&\\    
\hline    
5.0 &  -4.30 (320)       & -4.31 (1024)      & -4.26 (1024)    & -4.16 (1024)   \\
    &  -4.30 (980)	 &		     & -4.29 (3000)    & -4.26 (3544)\\
    &			 &		     &		       & -4.28 (14036) \\
\hline    
4.0 &  -1.30 (320)	 & -1.28 (1024)      & -1.26  (1024)   &\\
    &  -1.28 (500)	 &		     & -1.27  (3000)   &\\
\hline    
3.0 &  0.930 (500)	 &  0.943 (1024)     &  0.948 (1024)   &\\
    &			 &		     &  0.946 (3000)   &\\
\hline    
2.0 &  2.32  (500)	 &  2.35 (1024)      &  2.36 (1024)    &  2.33 (1024)\\
    &  2.35  (980)	 &		     &  2.35 (3000)    &  2.35 (3544)\\
    &			 &		     &		       &  2.36 (14036)\\
\hline    
1.0 &  3.32  (980)	 &  3.32 (1024)      &  3.23 (1024)    &  3.51 (1024)\\
    &			 &		     &  3.29 (3000)    &  3.36 (3544)\\
    &			 &		     &		       &  3.30 (14036)\\
\hline    
0.75&  3.13 (500)	 &  3.15 (1024)      &  3.02 (1024)    &  3.57 (1024)\\
    &  3.17 (980)	 &		     &  3.10 (3000)    &  3.26	(3544)\\
    &  3.16 (1620)	 &		     &		       &  3.19 (14036)\\
\hline
\hline
\end{tabular}
\end{center}
\label{table1}
\end{table}

\begin{table}
\caption{Comparison of energy $U$ and contact value $\rho_c$ of the
density profile for the two methods HK and EW3D. The surface charge
density is $\sigma=-1$ except for the case $h=12$ where $\sigma=-2$,
$\beta=1/kT$, $T$ temperature. The average density is
$\rho=N/V=  -2 \sigma / q h$.}

\vspace{1cm}
\begin{center}
\begin{tabular}{|ccc|cc|cc|}
\hline
\hline
\multicolumn{3}{|c|}{} & \multicolumn{2}{c|}{Ewald 3d} & \multicolumn{2}{c|}{Hautman-Klein}\\
\hline
&&&&&&\\    
$h$ & $q$ & $\rho$ & $\beta U/N$ & $\rho_c$  & $\beta U/N$ & $\rho_c$\\
&&&&&&\\    
\hline
&&&&&&\\    
5.0  & 5.0   & 0.08  & -4.30 & 6.30 & -4.31 & 6.38 \\
     & 4.0   & 0.1   & -1.28 & 6.30 & -1.28 & 6.32 \\
     & 3.0   & 2/15  & 0.93  & 6.30 & 0.94  & 6.31 \\
     & 2.5   & 0.16  & 1.70  & 6.30 & 1.71  & 6.29 \\
     & 2.0   & 0.2   & 2.35  & 6.30 & 2.35  & 6.29 \\
     & 1.5   & 4/15  & 2.97  & 6.35 & 2.98  & 6.33 \\
     & 1.0   & 0.4   & 3.32  & 6.53 & 3.32  & 6.54 \\
     & 0.75  & 8/15  & 3.17  & 7.22 & 3.15  & 7.20 \\  
     & 0.625 & 0.64  & 2.90  & 8.36 & 2.90  & 8.38 \\
     & 0.5   & 0.8   & 2.49  & 11.6 & 2.49  & 11.6 \\
&&&&&&\\    
8.0  & 1.0   & 0.25  & 3.46  & 6.40 & 3.53  & 6.39 \\
&&&&&&\\    
12.0$^{a}$ & 1.0  & 1/3 & 9.07 & 27.4 & 8.92  & 29.3\\
&&&&&&\\    
\hline
\hline
\multicolumn{7}{l}{${}^a \sigma=-2$}
\end{tabular}
\end{center}
\label{table2}
\end{table}


\newpage
\begin{thebibliography}{99}

%
\bibitem{Alle} Allen M P and Tildesley D J
1987 {\em Computer Simulation of Liquids} (Oxford:Clarendon) 

\bibitem{Sali} Salin G and Caillol J-M
2000 {\em J.\ Chem.\ Phys.\ } {\bf 113} 10459

\bibitem{Isra} Israelachvili J
1992 {\em Intermolecular \& Surface Forces} (San Diego:Academic) 

\bibitem{Evan} Evans D F and Wennerstr\"om H
1994 {\em The Colloidal Domain} (New York:VCH) 

\bibitem{Grun} von Gr\"unberg H H and Mbamala E C
2000 {\em J.\ Phys.\ : Condens. Matter} {\bf 12} 10349

\bibitem{Bell} De{'}Bell K, MacIsaac A B and Whitehead J P 
2000 {\em Rev.\ Mod.\ Phys.\ } {\bf 72} 225  

\bibitem{Weis} Weis J-J, Levesque D and Jorge S
2001 {\em Phys.\ Rev.\ } B {\bf 63} 045308  

\bibitem{Parr} Parry D E
1975 {\em Surf.\ Sci.\ } {\bf 49} 433

\bibitem{Parr1} Parry D E
1975 {\em Surf.\ Sci.\ } {\bf 54} 195

\bibitem{Heye} Heyes D M, Barber M and Clarke J H R 
1977 {\em J.\ Chem.\ Soc.\ Faraday Trans.\ II } {\bf 73} 1485

\bibitem{Leeu} de Leeuw S W and Perram J W 
1979 {\em Mol.\ Phys.\ } {\bf 37} 1313

\bibitem{Cich} Cichocki S W and Felderhof B U 
1989 {\em Mol.\ Phys.\ } {\bf 67} 1373

\bibitem{Harri} Harris F E 
1998  {\em Int.\ J.\ Quantum Chem.\ } {\bf 68} 385

\bibitem{Grzy} Grzybowski A, Gw\'o\'zd\'z and Br\'odka A 
2000 {\em Phys.\ Rev.\ B } {\bf 61} 6706

\bibitem{Spho} Spohr E 
1997 {\em J.\ Chem.\ Phys.\ } {\bf 107} 6342

\bibitem{Yeh} Yeh I-Ch  and Berkowitz M L 
1999 {\em J.\ Chem.\ Phys.\ } {\bf 111} 3155

\bibitem{Torr} Torrie G M and Valleau J P
1980 {\em J.\ Chem.\ Phys.\ } {\bf 73} 5807

\bibitem{Vall} Valleau J P, Ivkov R, and Torrie G M 
1991 {\em J.\ Chem.\ Phys.\ } {\bf 95} 520

\bibitem{Boda} Boda D, Chan  K-Y and Henderson D
1998 {\em J.\ Chem.\ Phys.\ } {\bf 109} 7362

\bibitem{Lekn} Lekner J
1989 {\em Physica } A {\bf 157} 826

\bibitem{Lekn1} Lekner J
1991 {\em Physica } A {\bf 176} 485

\bibitem{Maza} Mazars M 
2001 {\em J.\ Chem.\ Phys.\ } {\bf 115} 2955

\bibitem{Shel} Shelley J C and Patey G N 
1996 {\em Mol.\ Phys.\ } {\bf 88} 385

\bibitem{Croz} Crozier P S, Rowley R L, Spohr E and Henderson D 
2000 {\em J.\ Chem.\ Phys.\ } {\bf 112} 9253 

\bibitem{Haut} Hautman J and Klein M L 
1992 {\em Mol.\ Phys.\ } {\bf 75} 379

\bibitem{Cail} Caillol J-M and Levesque D 
1991 {\em J.\ Chem.\ Phys.\ } {\bf 94} 597

\bibitem{Pell} Pellenq R J-M, Caillol J-M and Delville A
1997 {\em J.\ Phys.\ Chem.\ } B {\bf 101} 8584  

\bibitem{Cail1} Caillol J-M 
1999 {\em J.\ Chem.\ Phys.\ } {\bf 111} 6528

\bibitem{Delv} Delville A, Pellenq R J-M and Caillol J-M 
1997 {\em J.\ Chem.\ Phys.\ } {\bf 106} 7275  

\bibitem{Alla} Allahyarov E, D {'}Amico I and L\"owen  H 
1999 {\em Phys.\ Rev.\ E } {\bf 60} 3199

\bibitem{Guld} Guldbrand P G, J\"onsson B, Wennerstr\"om H and Linse P 
1984 {\em J.\ Chem.\ Phys.\ } {\bf 80} 2221

\bibitem{Hend} Henderson D, Blum L and Lebowitz J L  
1979 {\em J.\ Electroanal.\ Chem.\ } {\bf 102} 315  

\bibitem{Carn} Carnie S L and Chan D Y C 
1981 {\em J.\ Chem.\ Phys.\ } {\bf 74} 1293  


\end{thebibliography}
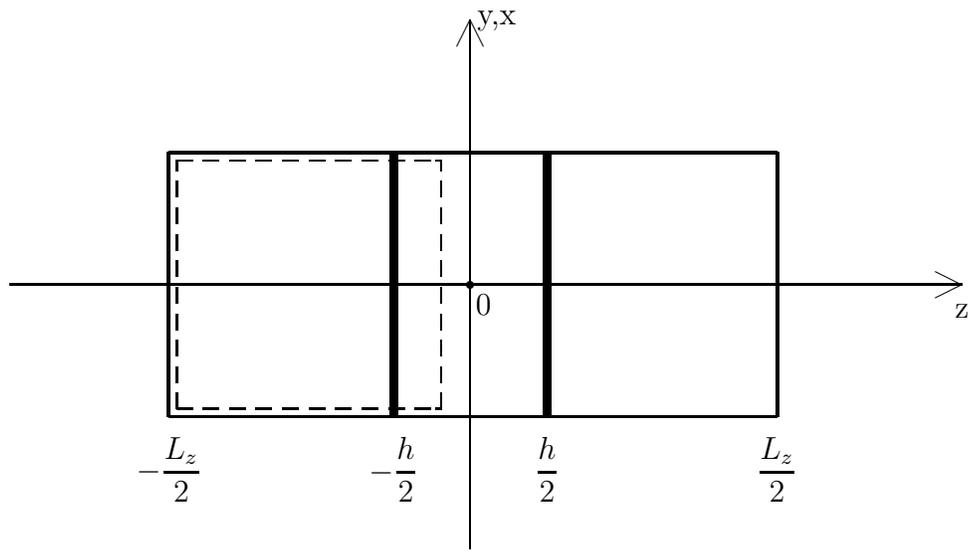
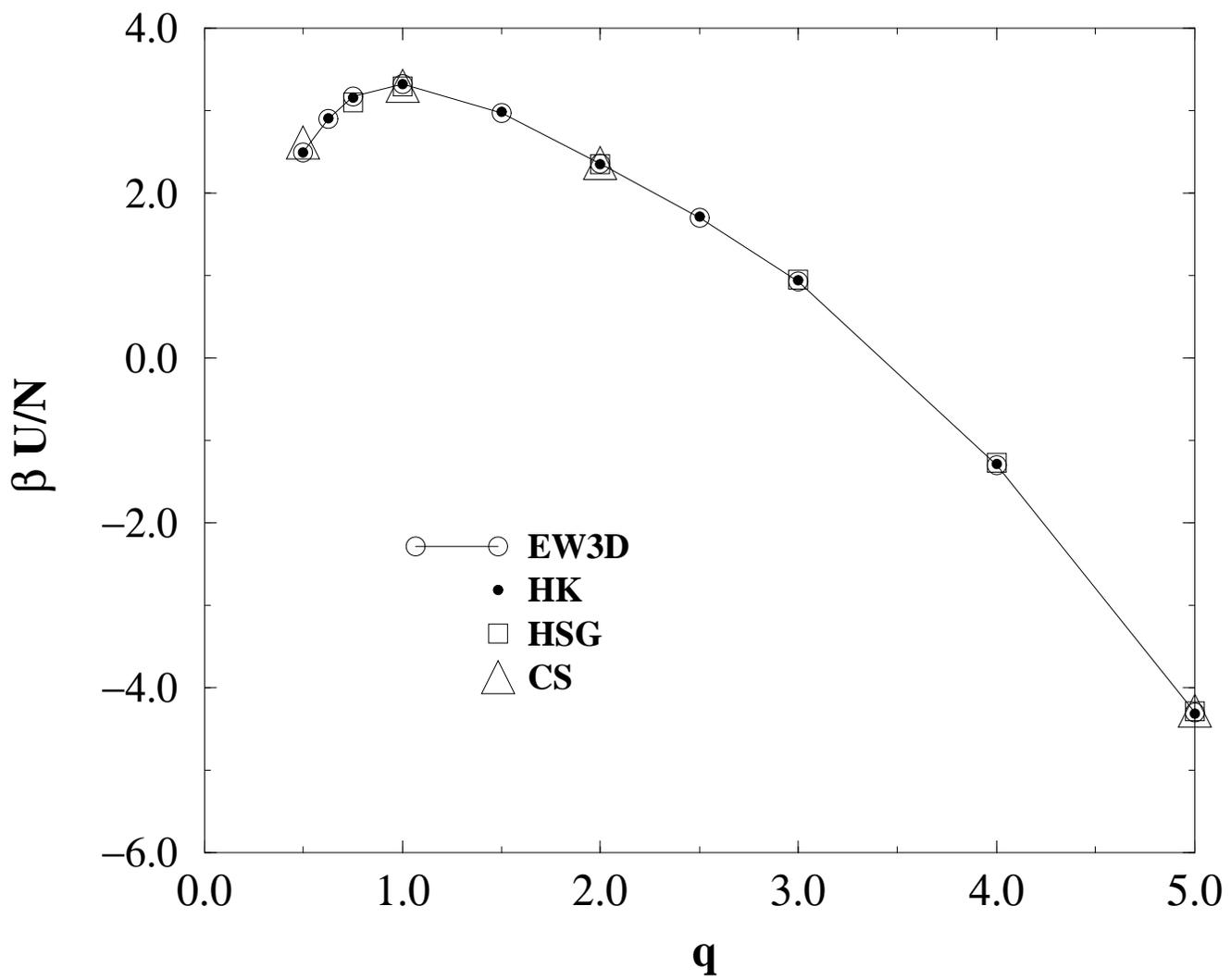
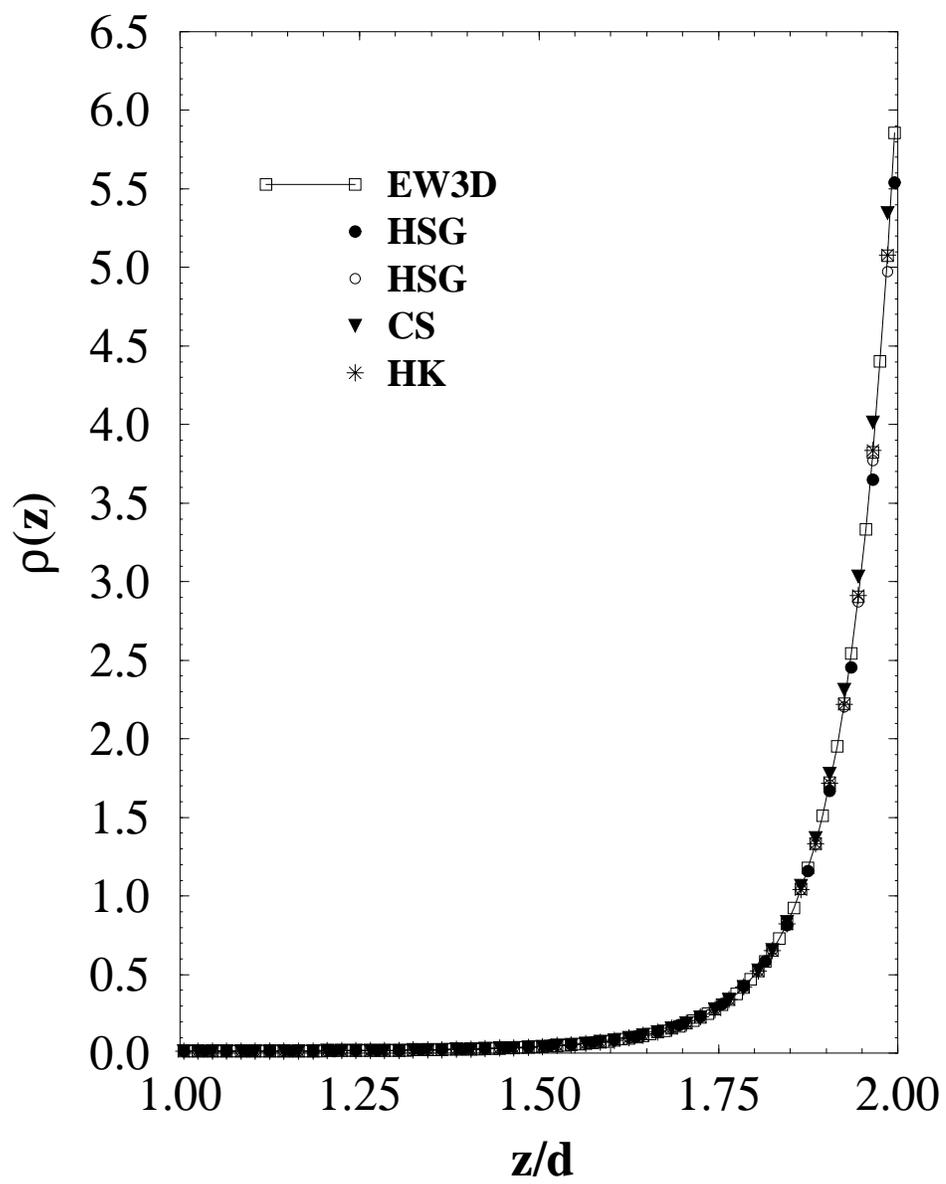
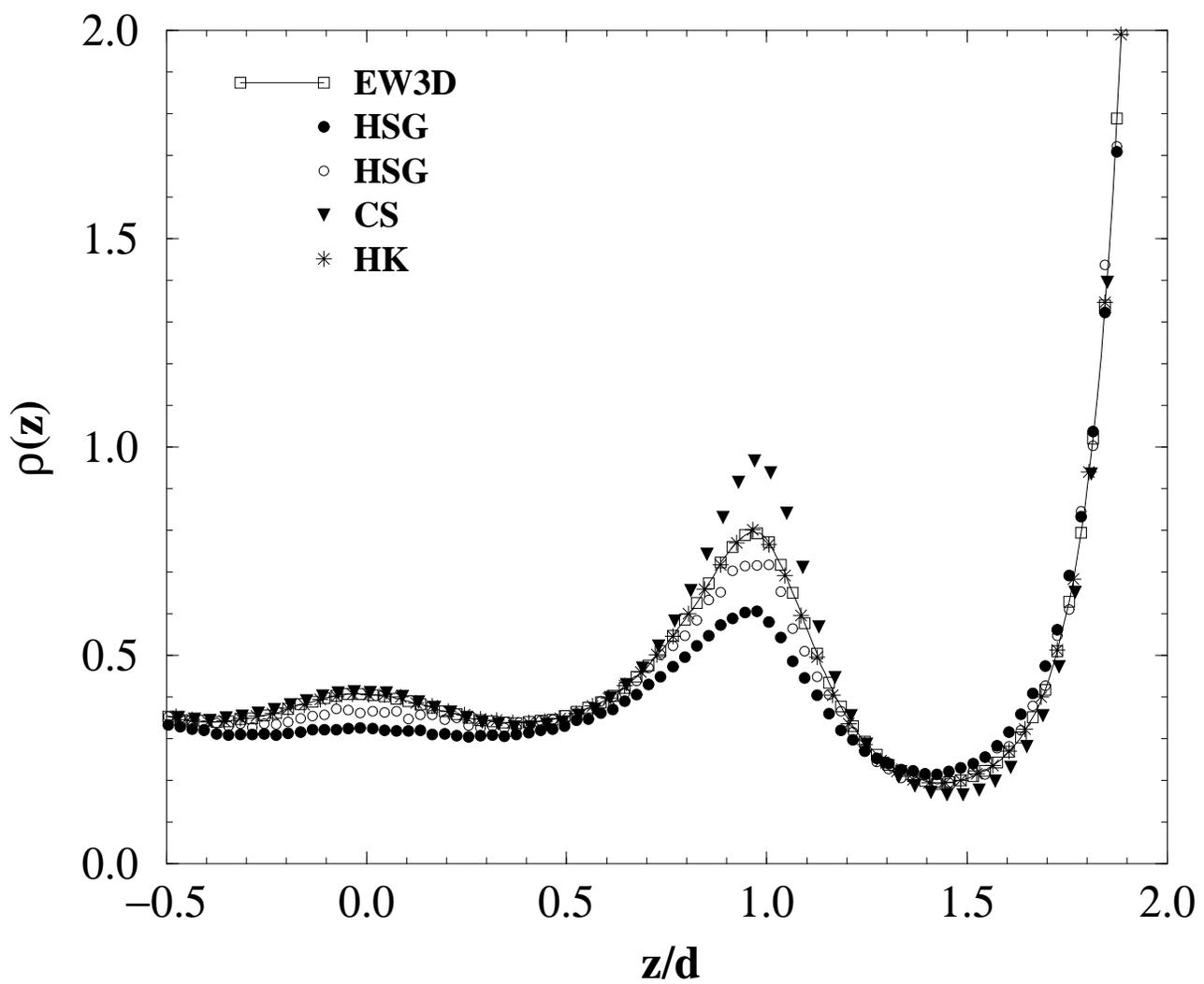
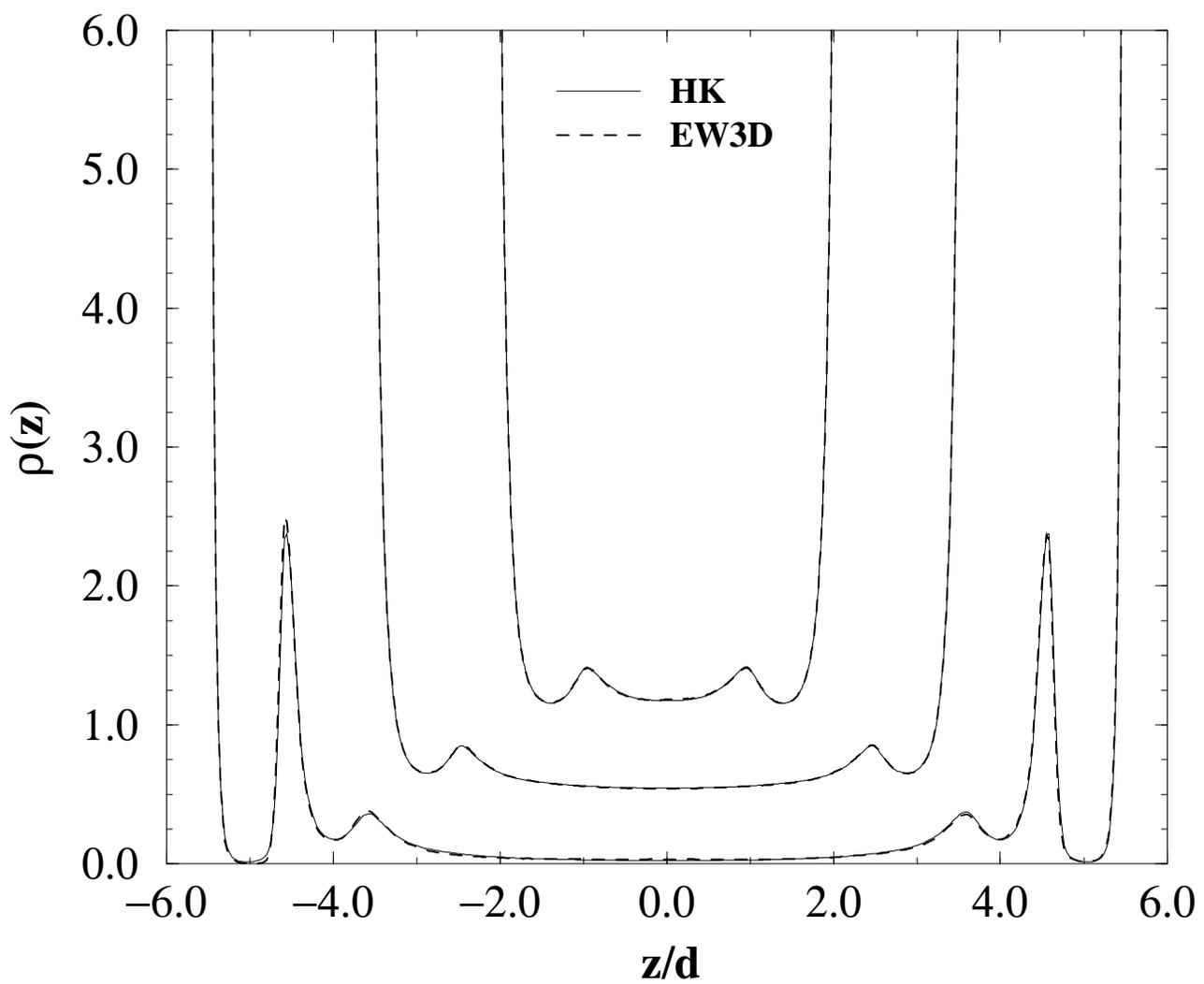
\end{document}